\documentclass[final,5p,times,twocolumn]{elsarticle}

\usepackage[utf8]{inputenc}
\usepackage{booktabs} 
\usepackage{lscape}
\usepackage{hyperref}
\usepackage{color}
\usepackage{soul}
\usepackage{booktabs}
\usepackage{longtable}
\usepackage{graphicx}
\usepackage{multirow}
\usepackage{multicol}
\usepackage{mathrsfs}
\usepackage{amssymb}
\usepackage{amsbsy}
\usepackage{amsmath}
\usepackage{tcolorbox}
\usepackage{balance}
\usepackage{microtype}
\usepackage{array}
\usepackage{longtable}
\usepackage{indentfirst}
\usepackage{pgfplots}
\usepackage{enumitem}
\usepackage{algorithmic}
\usepackage{textcomp}
\usepackage{xcolor}
\usepackage{eurosym} 
\usepackage[numbers]{natbib}

\newcommand{\newtext}[1]{{#1}}

\begin{document}

\title{\newtext{A Taxonomy for Mining and Classifying Privacy Requirements in Issue Reports}}


\author[1]{Pattaraporn Sangaroonsilp}
\ead{ps642@uowmail.edu.au}

\author[1]{Hoa Khanh Dam}
\ead{hoa@uow.edu.au}

\author[2]{Morakot Choetkiertikul}
\ead{morakot.cho@mahidol.ac.th}

\author[2]{Chaiyong Ragkhitwetsagul}
\ead{chaiyong.rag@mahidol.ac.th}

\author[1]{Aditya Ghose}
\ead{aditya@uow.edu.au}

\begin{abstract}
	
\textbf{Context:} Digital and physical trails of user activities are collected over the use of software applications and systems. As software becomes ubiquitous, protecting user privacy has become challenging. With the increase of user privacy awareness and advent of privacy regulations and policies, there is an emerging need to implement software systems that enhance the protection of personal data processing. However, existing data protection and privacy regulations provide key principles in high-level, making it difficult for software engineers to design and implement privacy-aware systems. \\
\textbf{Objective:} In this paper, we develop a taxonomy that provides a comprehensive set of privacy requirements based on four well-established personal data protection regulations and privacy frameworks, the General Data Protection Regulation (GDPR), ISO/IEC 29100, Thailand Personal Data Protection Act (Thailand PDPA) and Asia-Pacific Economic Cooperation (APEC) privacy framework. 
\textbf{Methods:} These requirements are extracted, refined and classified (using the goal-based requirements analysis method) into a level that can be used to map with issue reports. We have also performed a study on how two large open-source software projects (Google Chrome and Moodle) address the privacy requirements in our taxonomy through mining their issue reports. \\
\textbf{Results:} The paper discusses how the collected issues were classified, and presents the findings and insights generated from our study. \\
\textbf{Conclusion:} Mining and classifying privacy requirements in issue reports can help organisations be aware of their state of compliance by identifying privacy requirements that have not been addressed in their software projects. The taxonomy can also trace back to regulations, standards and frameworks that the software projects have not complied with based on the identified privacy requirements.


\end{abstract}


\maketitle

\section{Introduction} \label{sec:intro}


Software applications have become an integral part of our society. Digital trails are collected as people are browsing the Internet or using various software applications such as for social networking, working, studying and leisure activities. Physical trails are also collected through software systems such as surveillance cameras, face recognition apps, IoT sensors and GPS devices even when people are ``offline'' doing their normal life activities. Zettabytes of those data are collected and processed for various purposes \cite{Statista2021a}, including extracting and using personal data, and forming behavioural profiles of individuals. This poses serious threats to our privacy and the protection of our personal sphere of life -- the cornerstone of human rights and values.

Organisations have been collecting personal data of their customers for various business purposes. Cyberattacks often target at obtaining this data. CSO Online reported the fourteen biggest data breaches of the 21st century that affected 3.5 billion people \cite{Swinhoe2020}. The cases occurred with the world's top software applications, for examples, Adobe, Canva, eBay, LinkedIn and Yahoo.

The recent advent of privacy legislations, policies and standards (e.g. the European's General Data Protection Regulation \cite{OfficeJournaloftheEuropeanUnion;2016} or the ISO/IEC standard for privacy framework in information technology \cite{ISO/IEC2011}) aims to mitigate those threats of privacy invasion. A range of frameworks (e.g. privacy by design) and privacy engineering methodologies have also emerged to help design and develop software systems that provide acceptable levels of privacy and meet privacy regulations \cite{Ayala-Rivera2018}, \cite{Deng2011}, \cite{Aljeraisy2020}. However, those methodologies provide only high-level principles and guidelines, leaving a big gap for software engineers to fill in designing and implementing privacy-aware software systems \cite{Gurses2011}. Software engineers often face challenges when navigating through those regulations and policies to understand and implement them in software systems \cite{Ayala-Rivera2018}, \cite{Aljeraisy2020}, \cite{Senarath2018b}.


Hence, there is an emerging need to translate complex privacy concerns set out in regulations and standards into requirements that are to be implemented in software applications. Such privacy requirements need to be refined into a level that emphasises the functionalities in software systems and can be later used to map with issue reports. Previous work has involved extracting privacy requirements, but they are specific to a certain application domain such as e-commerce applications (e.g. \cite{Anton2002}) or healthcare websites (e.g. \cite{Antn2004}). More recent studies (e.g. \cite{Anthonysamy2017} and \cite{Guarda2009}) revealed an urgent need for a reference taxonomy of privacy requirements that are based on well-established regulations and standards such as GDPR and ISO/IEC 29100.

This taxonomy would be useful to understand how privacy requirements are explicitly addressed in software projects, and also serve as a basis for developing future privacy-aware software applications. Most of today's software projects follow an agile, issue-driven style in which feature requests, functionality implementations and all other project tasks are usually recorded as issues (e.g. JIRA issues\footnote{https://www.atlassian.com/software/jira}) \cite{Choetkiertikul}. In those projects, issue reports essentially contain important, albeit implicit, information about the requirements of a software, in the form of either new requirements (i.e. feature requests), change requests for existing requirements (i.e. improvements) or reporting requirements not being properly met (i.e. bugs) \cite{Choetkiertikul}, \cite{MoodleTracker}, \cite{MoodleFeature}. A software project consists of past issues that have been closed, ongoing issues that the team are working on, and new issues that have just been created. Through a study of those issues in the project, we can understand how the software team has implemented privacy requirements recorded in the issue tracking system (ITS) in order to address relevant privacy needs and concerns of stakeholders. This paper provides the following contributions:

\begin{enumerate}[leftmargin=0.5cm]
  \item We developed a comprehensive taxonomy of privacy requirements for software systems by extracting and refining requirements from the widely-adopted GDPR and ISO/IEC 29100 privacy framework as well as the newly developed Thailand PDPA and the region-specific APEC privacy framework. We followed a grounded theory process adapted from the Goal-Based Requirements Analysis Method (GBRAM) \cite{Antn2004} to develop this taxonomy. The taxonomy consists of 7 categories and 71 privacy requirement types. To the best of our knowledge, this taxonomy is well-grounded in standardised privacy regulations and frameworks as it covers more regulations and frameworks and the number of articles compared to the existing work (e.g. \cite{Meis}).  

  \item We mined the issue reports of two large-scale software projects, Chrome and Moodle (each has tens of thousands of issues) to extract 1,374 privacy-related issues. We classified all of those issues into the privacy requirements of our taxonomy. The classification was performed by multiple coders through multiple rounds of training sessions, inter-rater reliability assessments and disagreement resolution sessions. This resulted in a reliable dataset for the research community to perform future research in this timely, important topic of software engineering such as automated classification of privacy issue reports.

 \item We studied how the privacy requirements in our taxonomy were addressed in Chrome and Moodle issue reports. We found \newtext{2,432} occurrences of the privacy requirements in our taxonomy were covered in those datasets (see Section \ref{sec:results} for more details), most of which related to the user participation category. In addition, we found that allowing the erasure of personal data is a top concern reported in the Chrome and Moodle issue reports, while none of the privacy requirements related to the management perspective of data controllers was recorded in the issue tracking system of both projects. We also discovered that privacy and non-privacy issues were treated differently in terms of resolution time and developers' engagement in both projects.


\end{enumerate}

A full replication package containing all the artifacts and datasets produced by our studies are made publicly available at \cite{reppkg-pridp}. The remainder of this paper is structured as follows. Section \ref{sec:related-work} provides related existing work on privacy requirements engineering. \newtext{A theoretical background of the privacy requirements taxonomy is discussed in Section \ref{sec:theoretical-background}. The methodology used to build the taxonomy is presented Section \ref{sec:taxonomy-development}. Section \ref{sec:taxonomy} describes the privacy requirements taxonomy and its categories in detail. Section \ref{sec:mining} presents our study of how Chrome and Moodle issue reports address the privacy requirements in our taxonomy. The findings and insights generated from the study are discussed in Section \ref{sec:results}. Section \ref{sec:threats} discusses the threats to validity. Finally, we conclude and discuss future work in Section \ref{sec:conclusion}.}

\section{Related work} \label{sec:related-work}

The protection of personal data and privacy of users have attracted significant attention in recent years. Data protection regulations and privacy frameworks have been established to guide the development of software and information systems. \emph{However, it is challenging for software engineers to translate legal statements stated in those regulations into specific privacy requirements for software systems \cite{Ayala-Rivera2018}, \cite{Aljeraisy2020}}. Data protection and privacy laws are independently designed and enforced for specific areas, which could range from a state (e.g. California), a country (e.g. Australia) or a region (e.g. Europe and Asia Pacific). Any organisations that meet the conditions of these laws must comply. However, the developers lack guidance and also experience difficulties in understanding and extracting such privacy requirements from those required laws \cite{Ayala-Rivera2018}, \cite{Aljeraisy2020}. Several studies have been calling for frameworks and methodologies to support software engineers in designing and developing privacy-aware software systems \cite{Gurses2011}, \cite{Senarath2018b}, \cite{Sheth2014}, \cite{Birnhack2014}. 

\citeauthor{Beckers2012} \cite{Beckers2012} proposed a conceptual framework to compare privacy requirements engineering approaches on requirements elicitation and notion representation. The approaches include LINDDUN, PriS and the framework for privacy-friendly system design approaches. The LINDDUN method elicits privacy requirements by modeling a system using a Data Flow Diagram (DFD) \cite{Deng2011}. The elements in the DFD are then mapped to the privacy threat categories to identify privacy requirements. \citeauthor{Kalloniatis2008} \cite{Kalloniatis2008} proposed a PriS method to elicit privacy requirements in the software design process. Privacy requirements in this study are modelled as a type of organisational goals that needs to be achieved in a specific application. Comparing with our study, LINDDUN and PriS methods do not elicit requirements from privacy and data protection regulations and frameworks, and they employed different requirements elicitation approaches.

Several existing work have aimed to identify privacy requirements and construct a privacy requirement taxonomy from privacy policies, regulations and standards. Ant\'{o}n et al. \cite{Anton2002}, \cite{Antn2004} used the GBRAM to develop the taxonomy of privacy requirements from the privacy policies of e-commerce and health care websites. The taxonomy was constructed by applying goal identification and refinement strategies to extract goals and requirements. We adapted this approach to construct the taxonomy presented in this paper. \citeauthor{Meis} \cite{Meis}, \cite{Meis2016} proposed a taxonomy of transparency requirements to support software engineers in identifying relevant requirements from a draft version of GDPR and ISO/IEC 29100. However, these studies emphasise on privacy goal transparency and intervenability in software development. \citeauthor{Gharib2017} \cite{Gharib2017} proposed an ontology that identifies key concepts for capturing privacy requirements. However, these concepts provide high-level dimensions rather than software requirements level. \citeauthor{Ayala-Rivera2018} \cite{Ayala-Rivera2018} proposed an approach to map GDPR data protection obligations with privacy controls derived from ISO/IEC standards. Those links help elicit the solution requirements that should be implemented in a software application. However, their study focused and validated only two articles in GDPR (i.e. Articles 5 and 25).

\newtext{The following work addressed the methods to elicit privacy requirements from software artifacts and patterns of GDPR requirements. \citeauthor{Colesky} introduced tactics which were used to link between privacy design strategies and privacy patterns \cite{Colesky}. The tactics can be considered as brief requirements which provide a guide to achieve privacy protection based on the privacy design strategies. This study associated those strategies, tactics and patterns to some GDPR entities and personal data processing examples. However, the tactics were defined in high-level and covered only one GDPR article.}

\newtext{\citeauthor{Ferreyra2020} proposed a method named PDP-ReqLite to elicit privacy and data protection requirements in systems and software projects \cite{Ferreyra2020}. It received Requirements DFD and Personal Information Diagram (PID) as inputs and generated meta-requirements in the form of pre-condition and post-condition predicates. The meta-requirements were patterns derived from translating the statements in GDPR directives and principles. Those meta-requirements were later combined into the respective GDPR categories. The study claimed that it ensured the full coverage of GDPR directives, however it only demonstrated the elicitation of undetectability requirements and did not clearly specify which GDPR articles were covered.}

\newtext{\citeauthor{Notario2015} developed a systematic methodology that combined risk-based and goal-oriented approaches to transform high-level privacy principles into operational privacy requirements \cite{Notario2015}. It identified relevant privacy principles based on organisational goals and/or regulatory frameworks, determined the required level of conformance for a system and determined the applicability of each requirement depending on other system and organisational constraints. However, the study only addressed the accountability principle in GDPR, and the method focused on software analysis and design processes.}

Several researchers proposed models and heuristics to represent and extract requirements from regulations. \citeauthor{Breaux2006} \cite{Breaux2006} proposed a process called Semantic Parameterisation to extract rights and obligations from the Privacy Rule from the U.S. Health Insurance Portability and Accountability Act (HIPAA). \citeauthor{Breaux2013} \cite{Breaux2013} developed a legal requirements specifications language (LRSL) to codify policy and law requirements from thirteen U.S. state data breach notification laws. This method also supports the traceability of regulatory requirements across multiple jurisdictions. However, these methods focused on specific schemes (i.e. HIPAA and breach notification).

Several work developed a tool support to automatically extract requirements from legal documents. \citeauthor{Zeni2015} \cite{Zeni2015} uses the semantic annotation (SA) technique to extract rights and obligations from HIPAA and Italian accessibility law for information technology instruments. The tool can capture the hierarchical structure and cross-references of legal documents and support the annotation of different languages other than English. \citeauthor{Sleimi2018} \cite{Sleimi2018} identified the metadata types of legal requirements from traffic laws and annotated the legal statements with the identified metadata types to generate NLP-based rules. Those rules were then implemented to automatically extract metadata types from the traffic laws of Luxembourg. Later, the same group of authors proposed a homonised set of legal requirements templates that systematically expresses legal requirements from multiple viewpoints \cite{Sleimi2020}. They also developed a tool support to automatically recommend those templates for the statements in the Luxembourg labour and health laws.

The following work discusses privacy requirements extraction from regulations for compliance checking in software systems. \citeauthor{Torre} \cite{Torre} developed a conceptual model using hypothesis coding to specify metadata types that exist in the statements of selected GDPR articles and created dependencies between those metadata types to ensure the proper completeness checking. They also employed Natural Language Processing (NLP) and Machine Learning (ML) techniques to automatically extract and classify the metadata in privacy policies from the fund domain. \citeauthor{Ghanavati2009} \cite{Ghanavati2009} developed a framework to analyse to what degree the organisation complies with laws. The study adopted a Goal-oriented Requirements Language (GRL) to model and analyse the relationship between organisational and legal requirements, and identify which organisational goals satisfy requirements in the law. This method was evaluated on Ontario's Personal Health Information Protection Act (PHIPA) only. 

Particularly focusing on the work related to GDPR, the European Commission has funded the GDPR cluster projects to help tackle the GDPR implementation challenges faced by organisations (e.g. \cite{BPR4GDPR}, \cite{DEFEND}, \cite{SMOOTH}, \cite{PDP4E}, \cite{PAPAYA}, \cite{POSEIDON} and \cite{Gharib2016a}). Those projects have developed both organisational and technical techniques to facilitate the implementation. They have addressed different challenges complying with GDPR in software development activities (e.g. planning, design, development, operation and deployment). They also provide solutions to the identified challenges. \citeauthor{EUcluster2020} \cite{EUcluster2020} has summarised the solutions proposed by some of these projects.

\section{\newtext{Theoretical Background}} \label{sec:theoretical-background}

Many countries around the world have been developing data protection and privacy legislation to strengthen their personal data and privacy protection \cite{UNCTAD2020}. These legislations are designed to provide organisations with a comprehensive benchmark to govern their personal data collection and processing as well as protect and empower individuals about their privacy and rights. Having the legislations in place seems to benefit both organisations and individuals, however many organisations have faced several challenges to comply with these legislations \cite{Capgemini2019}. Those challenges raise the need to develop a taxonomy of requirements from data protection and privacy regulations to support the development and compliance of privacy-aware software systems.


Our work is based on two well-established and widely-adopted regulations and privacy frameworks: GDPR and ISO/IEC 29100 and two region-specific representatives: Thailand PDPA and APEC privacy framework. GDPR is enacted to protect the individual rights of the data subjects on their personal data \cite{OfficeJournaloftheEuropeanUnion;2016}. It provides conditions, principles and definitions that need to be integrated into organisational processes and policies. These processes involve the collection, use, process, storage and dissemination of personal data of EU citizens and residents. The organisations failing to comply with the GDPR can be fined up to \EUR{20} million or 4\% of their previous year’s global turnover, whichever is greater. After a year of the enforcement, there are over 230 finalised cases with the total of \EUR{150} million fines so far. A great number of GDPR violation cases related to the processing of personal data and data breach have been reported \cite{EuropeanCommission2019}, \cite{PrivacyAffa}. This suggests the challenges in operationalising GDPR in developing software applications.

ISO/IEC 29100:2011 is a privacy framework which guides the processing of personally identifiable information (PII) in Information and Communication Technology systems \cite{ISO/IEC2011}. The framework defines a set of privacy principles used to handle personal data processing activities (e.g. collection, storage, use, transfer and disposal). \emph{Similarly to GDPR, those principles are high-level, making it challenging for software engineers to design and implement privacy-aware systems}. We aim to address these challenges by translating these complex statements into implementable privacy requirements for software systems.

Thailand Personal Data Protection Act (PDPA) was officially announced in May 2019 \cite{PDPA}. The regulation came into full effect in June 2022 after several extensions. Thailand PDPA is designed to govern personal data protection and create transparency and fairness for the use of personal data. It also promotes the use of personal data for innovation under assurance and provides effective remedy from data breaches. Any organisations that collect, use and disclose personal data of individuals residing in Thailand must comply with the regulation. We include this regulation in our study as it is a representative of newly developed and country-specific personal data protection regulation.


Asia-Pacific Economic Cooperation (APEC) privacy framework 2015 \cite{Apec2015} was published in August 2017 with the intention to establish effective privacy protections for cross-border information transfers across member countries of APEC \footnote{A list of APEC member countries can be found at https://www.apec.org/about-us/about-apec}. The APEC privacy provides guidance and direction to businesses and government entities in APEC economies on developing appropriate privacy protections of all personal information to ensure the free flow information in the Asia Pacific region. Unlike the GDPR, the framework does not displace domestic laws of the member countries. Also, it is not generally followed outside this region, except for outsider providing services to this region. Thus, it is included in this study as a representative of region-specific privacy protection framework.

\newtext{We selected GDPR and ISO/IEC 29100 since they attracted a lot of attention from the public and also in the literature. In addition, the enactment of GDPR has greatly affected the way how organisations handle personal data in many sectors around the world. The data subjects also better aware of their individual rights in managing their personal data from GDPR. Both GDPR and ISO/IEC 29100 focus on data subjects and their personal data which can be applied to any sectors unlike, for example, California Consumer Privacy Act (CCPA) \cite{StateofCaliforniaDepartmentofJustice2018} which specifically covers consumers in businesses. Thailand PDPA and APEC privacy framework have been selected as representatives to validate the commonalities of privacy requirements across different regions.}

\newtext{Prior to this work, we have conducted a thorough study on GDPR, ISO/IEC 29100, Thailand PDPA and APEC privacy framework and have found that they share many commonalities\footnote{See the file \emph{Privacy-requirements-abstraction} in the replication package for some examples \cite{reppkg-pridp}.}.} All of the regulations, standards and frameworks provide benchmarks for privacy and data protection governance and compliance in organisations. They are common in laying out the expectations that should be met when handling personal data. They also complement each other to cover a comprehensive set of privacy-related software requirements. We have done the mapping between the GDPR, ISO/IEC 29100, Thailand PDPA and APEC framework principles to demonstrate their similarities\footnote{See the file \emph{Mapping-across-regulations} in the replication package for more details \cite{reppkg-pridp}.}.  

\section{\newtext{Privacy Requirements Taxonomy Development}} \label{sec:taxonomy-development}

\begin{figure}
	\centering
	\includegraphics[width=1\linewidth]{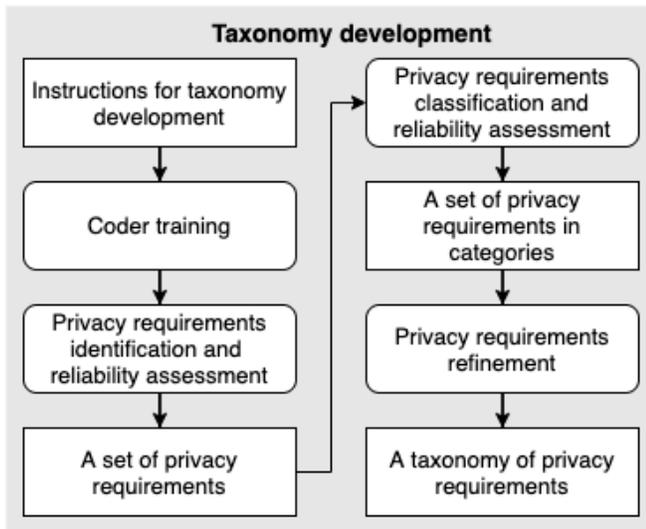}
	\caption{\newtext{An overview process of privacy requirements taxonomy development}}
	\label{fig:taxonomy development}
\end{figure}

This section discusses the methodology that we have followed to develop this taxonomy  (see Figure \ref{fig:taxonomy development}). We followed a content analysis process adapted from the GBRAM \cite{Antn2004}, which is based on Grounded Theory, to develop a taxonomy of privacy requirements. GBRAM is a systematic method used to identify, refine and organise goals into software requirements. This process was applied to analyse goals from natural language texts in privacy policies, and convert them into software requirements. The method has been successfully applied to the analysis of e-commerce applications \cite{Anton1998} and health care privacy policies \cite{Antn2004}. The method consists of three main activities: goal identification, goal classification and goal refinement. Goal identification derives goals from specifications in selected sources. Each identified goal is then classified into one of the pre-defined categories in the goal classification. Finally, the goal refinement removes synonymous and redundant goals, resolves inconsistencies among the goals and operationalises the goals into requirements specification. 

We had multiple researchers (co-authors of the paper, hereby referred to as the coders) follow this process to develop the taxonomy independently, and used the inter-rater reliability (IRR) assessment to validate the agreements and resolve disagreements. Those coders were given instructions and trained at the start of the process. The process consists of the following steps:

\begin{itemize}[leftmargin=*, noitemsep]
	\item \textbf{\newtext{Step 1} - privacy requirements identification:} extract requirements from written statements in the studied privacy regulations and frameworks, and structure them into a pattern (action verb, objects and object complement).
	
	\item \textbf{\newtext{Step 2} - privacy requirements refinement:} remove duplicate requirements and manage inconsistent requirements. Since the inputs were written in descriptive statements and from different sources, requirements can be redundant or inconsistent.
	
	\item \textbf{\newtext{Step 3} - privacy requirements classification:} classify requirements into categories based on a set of privacy goals. The privacy goals can be considered as a group of functionalities that the software systems are expected to provide.
\end{itemize}

The details of each step are described as follows.
\newtext{\subsection{Privacy requirements identification}} \label{subsec:req-identification}

This step aims to identify privacy requirements from the narrative statements in GDPR, ISO/IEC 29100 privacy framework, Thailand PDPA and APEC privacy framework. We first created a range of questions to identify goals from each statement of the studied privacy regulations and frameworks. As part of our research, we have carefully manually gone through all 99 articles in the GDPR. Although GDPR and Thailand PDPA govern broader regulatory aspects comparing to ISO/IEC 29100 and APEC privacy frameworks such as requirements in roles assignment (e.g. data protection officer and supervisory authority), managing juridical remedies and noticing penalties, those are not software requirements. Thus, they are out of the scope of our study. 

We then applied several filters to select the articles that address software requirements in GDPR and include them in our study. We selected 19 articles that address the rights of individuals and cover the key principles of the GDPR (i.e. Articles 6-7, 12-22, 25, 29-30, 32-34). However, we did not consider Chapter V (Articles 44 - 50) since it focuses on legal administrative perspective for international border transfer rather than the software application level. Thus, we did not analyse them as software requirements in our study. For the ISO/IEC 29100 privacy framework, all of the contents were explored.

Thailand PDPA consists of 7 chapters, 96 articles and 7 rights of data subjects. We have analysed 16 articles in Chapter 2 (Personal data protection) and Chapter 3 (Rights of data subjects) which are the key chapters providing guidelines to govern personal data and privacy protection. Other chapters detailing the scope of use, definitions, assignment of personal data protection committee and supervisory authority, complaints and penalties are not included in the taxonomy development process as they are not related to software requirements. In the APEC privacy framework, there are four parts containing 72 points. We have analysed the APEC information privacy principles part as it is related to software requirements. Based on the scope defined above, we shortlisted 149 statements in GDPR, 63 in ISO/IEC 29100, 101 in Thailand PDPA and 74 in the APEC privacy framework to be explored. This initial set of statements was directly extracted from the list of itemised items and/or clauses in the selected parts of regulations and frameworks.

Next, we went through all the shortlisted statements. We analysed each statement using a set of pre-defined questions to identify relevant actions, involved/affected parties or objects and target results. A statement may cover more than one requirement.  The steps of privacy requirements identification process are explained below:


\begin{enumerate}[leftmargin=*, noitemsep]
	
	\item \textbf{Identifying actions:} We ask \textit{``Which action should be provided based on this statement?''} to identify the action associated with a requirement. Some examples of the action verbs used in the collected statements are: \textbf{ALLOW, COLLECT, ERASE, IMPLEMENT, INFORM, MAINTAIN, NOTIFY, OBTAIN, PRESENT, PROTECT, PROVIDE, REQUEST, SHOW, STORE, TRANSMIT and USE}.
	
	\item \textbf{Determining involved/affected parties or objects:} After an action is identified, we determine the object(s) of the action. The output from this step can be either involved/affected parties or objects that are directly identified or implied by the statements. The involved/affected parties can be any persons or stakeholders such as data subjects, data processors, data recipients, supervisory authorities or third parties. The question used to identify the involved/affected parties is ``Who is involved/affected from the statement?''. However, the objects are things that are created, processed or done by the actions specified in the statements (e.g. consent, preferences, personal data, functions and data repository). These objects are identified by asking ``What has to be created/done from the identified action?''. 
	
	\item \textbf{Considering the target result(s):} The target results refer to a goal that a statement aims to achieve. They can be identified by asking \textit{``What should be achieved based on the action of that statement?''} For example, Article 13(1)(c) in GDPR states ``..., the controller shall, at the time when personal data are obtained, provide the data subject with the purposes of the processing for which the personal data are intended as well as the legal basis for the processing;''. The goal in this statement is asking to provide the data subject with the purposes of processing. Hence, the purposes of the processing is a target result that the action verb \textit{PROVIDE} aims to achieve.
	
	\item \textbf{Structuring into a privacy requirement pattern:} The derived privacy requirement is coded in the format of action verb, followed by involved/affected parties or objects and target results.
	
\end{enumerate}


The following examples illustrate these steps. A statement in the GDPR states ``... the controller shall, at the time when personal data are obtained, provide the data subject the identity and the contact details of the controller and, where applicable, of the controller's representative''. From this statement, we identify \textit{`PROVIDE'} as the action that the controller shall act. We then consider what should be provided by the controller, and that was \textit{`the identity and the contact details of the controller or the controller's representative'}. We determine the object responding to \textit{`to whom the identity and contact details of the controller or the controller's representative should be provided'}, and that is the data subjects. All three components formulate a privacy requirement as \textit{\textbf{PROVIDE} the data subjects with the identity and contact details of a controller/controller's representative (R22)}.

Another example is more complex than the previous one. In the GDPR, removing personal data is recommended in several ways such as: the data has been unlawfully processed; or the data subjects would like to erase their personal data themselves; or the system must erase personal data when the data subjects object to the processing; or the system must erase personal data when the data subjects withdraw consent; or the system must erase personal data when it is not necessary for the specified purpose(s); or the system must erase personal data when the purpose(s) for the processing has expired. We thus need to formulate different privacy requirements as they affect the ways that the functions would be provided to users in a system. In this example, the derived requirements are: \textit{\textbf{ERASE} the personal data when it has been unlawfully processed (R7)}, \textit{\textbf{ALLOW} the data subjects to erase his/her personal data (R44)}, \textit{\textbf{ERASE} the personal data when the data subjects object to the processing (R46)}, \textit{\textbf{ERASE} the personal data when a consent is withdrawn (R47)}, \textit{\textbf{ERASE} the personal data when it is no longer necessary for the specified purpose(s) (R52)} and \textit{\textbf{ERASE} the personal data when the purpose(s) for the processing has expired (R53)}. More examples of the privacy requirements derived in this step are included in the supplementary material.

\textbf{Reliability assessment}: Three human coders, who are the co-authors of the paper, have independently followed the above process to identify privacy requirements from the GDPR and ISO/IEC 29100 privacy framework. All three coders had substantial software engineering background and at least 1 year of experience with data protection regulations and policies. The first author prepared the materials and detailed instructions\footnote{These are included in the replication package \cite{reppkg-pridp}.} for the process. The instructions were provided to all the coders before they started the identification process. A 1-hour training session was also held to explain the process of identifying privacy requirements, clarify ambiguities and define expected outputs. The coders also went through a few examples together to fine tune the understanding.

All the coders were provided with a form to record their results of each step. The form was pre-filled with 149 statements extracted from the GDPR and 63 statements from the ISO/IEC 29100 privacy framework. If a coder considers a statement as a privacy requirement, they need to identify the relevant components and structure it following the privacy requirement patterns above. Otherwise, they leave it blank. Initially, the three coders each respectively identified 100, 95 and 97 requirements from GDPR, and 36, 36 and 37 requirements from ISO/IEC 29100.

Since the requirements identified by the coders could be different, we used the Kappa statistic (also known as Kappa coefficient) to measure the IRR between the coders \cite{Viera2005}. The Kappa statistic ranges from -1 to 1, where 1 is perfect agreement and -1 is strong disagreement \cite{Viera2005}. There are several types of the Kappa statistics which suit different study settings \cite{Hallgren}. For this study, the Fleiss' Kappa was used as we had three coders coding the same datasets \cite{Fleiss1971}. The Kappa values were 0.8025 for GDPR and 0.7182 for ISO/IEC 29100, suggesting a substantial agreement level amongst all the coders \cite{Landis1977}. All the coders agreed that there were 43 and 20 statements from the GDPR and ISO/IEC 29100 respectively that are not privacy requirements. There were 20 GDPR statements (and 13 for ISO/IEC 29100) that the coders did not agree upon. Hence, a meeting session was held between the coders to discuss and resolve disagreements. 

The statements in Thailand PDPA and APEC privacy framework were identified by one of the coders using the same methodology performed with the GDPR and ISO/IEC 29100. We used only one coder because the provisions in Thailand PDPA and many principles in the APEC privacy framework share many commonalities with the GDPR and ISO/IEC 29100, respectively. We therefore decided that one coder would be sufficient. The coder, who was responsible for the Thailand PDPA and APEC privacy framework, was the main coordinator and also participated in the privacy requirements identification process for GDPR and ISO/IEC 29100. All the shortlisted statements in Thailand PDPA and APEC privacy framework were derived in this step. This brought the total number of privacy requirements obtained in this step to 249 (116 from GDPR, 33 from ISO/IEC 29100 and 55 from Thailand PDPA and 45 APEC privacy framework).



\newtext{\subsection{Privacy requirements refinement}} \label{subsec:req-refinement}

Requirements extracted either from the same or different documents can be similar, redundant or inconsistent. In this step, we identify those similar and duplicate requirements, and merge them into one single requirement. In case that the requirements are inconsistent, we perform further investigation and report for notice.

To identify and merge similar requirements, we first place those similar requirements into the same group. These requirements tend to achieve the same goal and have the same involved/affected parties or objects. We then determine the action and target result for the final merged requirement based on the following rules:
\begin{enumerate}[leftmargin=*, noitemsep]
	\item If the action verbs in the requirements are the same, we retain that action for the final merged requirement.
	\item If the actions are different, we consider the action verb based on the following:
	\begin{itemize}[leftmargin=*, noitemsep]
		\item Use \emph{ALLOW} if a requirement relates to data subject's ability to invoke his/her rights.
		\item Use \emph{PROVIDE} if a requirement aims to give information to stakeholders.
		\item Use \emph{OBTAIN} if a requirement aims to get a consent or permission from stakeholders.
		\item Use \emph{PRESENT} if a requirement aims to display options or choices to stakeholders. This action verb requires responses from the stakeholders (e.g. displaying toggles or radio buttons for users to select).
		\item Use \emph{SHOW} if a requirement aims to show information to stakeholders. This action verb does not require any responses from the stakeholders.
		\item Use \emph{NOTIFY} if a requirement aims to alert stakeholders.
		\item Use \emph{IMPLEMENT} if a requirement aims to build a mechanism to support an activity in a system.
		\item Use \emph{ERASE} if a requirement aims to erase data in software systems.
	\end{itemize}
	\item If the target results in the requirements are the same, we retain that target result for the final requirement.
	\item If the target results are different, we combine them together. In case they have redundant or synonymous words, we select one word from the words in the list.
\end{enumerate}	

We finally put together the action, involved/affected parties or objects and target results identified in the above steps to construct the final requirement. 

We have carefully investigated the terms and definitions used in GDPR and ISO/IEC 29100 and found that they mostly refer to similar or same things. For example, ISO/IEC 29100 defines PII as ``any information that (a) can be used to identify the PII principal to whom such information relates, or (b) is or might be directly linked to a principal''. Personal data in GDPR is defined as ``any information relating to an identified or identifiable natural person ('data subject')''. As can be seen, PII in ISO/IEC 29100 and personal data in GDPR in fact refer to the same thing, i.e. any information that can be used to identify or is linkable to a natural person. Similarly, a PII principal in ISO/IEC 29100 and a data subject in GDPR in fact refer to the same thing, i.e. a natural person who can be identified with identifiable information such as name.

Similarly, key terms GDPR and ISO/IEC 29100 are sometimes worded differently, however their definitions are similar. For example, \textit{`processing'} means ``any operation or set of operations which is performed on personal data or set of personal data, ...'' in GDPR and \textit{`processing of PII'} is defined as ``operation or set of operations performed upon personally identifiable information (PII)''. Another example is data controller in GDPR and PII controller in ISO/IEC 29100. Both terms refer to person that determines the purposes and means of the processing of personal data/PII. Other examples include personal data with PII, data subject with PII principal, data processor with PII processor, consent and third party. Thus, in the merging step, we use the terms from GDPR in representing our requirements in this taxonomy to avoid ambiguities. We have also found that privacy requirements derived from GDPR, ISO/IEC 29100, Thailand PDPA and APEC privacy framework are in fact at the same level of abstraction\footnote{see the supplementary material for more details.}.

The following example demonstrates the requirements merging step. A statement in ISO/IEC 29100, ``... allow a PII principal to withdraw consent easily and free of charge ...'', derives a requirement \textit{\textbf{ALLOW} a PII principal to withdraw consent}. A statement in GDPR, ``... the controller shall ... provide the data subjects with ... the existence of the right to withdraw consent at any time ...'' gives a requirement \textit{\textbf{PROVIDE} the existence of the right to withdraw consent}. A statement in Thailand PDPA, ``The data subject may withdraw his or her consent at any time.'', gives a requirement \textit{\textbf{ALLOW} the data subject to withdraw his or her consent.} The goal of these three requirements is to let the PII principal/data subject withdraw consent, and the affected parties are PII principal and data subject. It is noted that we use the terms from GDPR for roles in our requirements (i.e. data subject, data controller, data processor and third parties). We therefore list them as similar requirements. The requirements have different actions (i.e. ALLOW and PROVIDE), we then use \emph{ALLOW} as the final action as these requirements are about data user's ability to withdraw consent. We acquire \textit{withdraw consent} as a common target result. Finally, we merge these three requirements into a single requirement: \textit{\textbf{ALLOW} the data subjects to withdraw consent (R6)}.

The duplicate requirements are the requirements that have the exact actions, involved/affected parties and target results. We represent these requirements as one requirement in the taxonomy. For example, we identify two exact requirements in the identification process, \emph{PROVIDE the data subject the categories of personal data concerned} in GDPR Art. 14(b) and 15(b). We retain one requirement (i.e. R42) in the taxonomy.

Requirements are inconsistent when they appear to contradict each other in performing the same actions. The following example demonstrates the consistency between the requirements in ISO/IEC 29100 and GDPR. We identify the requirements from ISO/IEC 29100 and GDPR as ``COLLECT only necessary PII for specific purposes'' and ``COLLECT the personal data as necessary for specific purposes'', respectively. Both requirements yield that the personal data must be collected as necessary for specific purposes. They are presented in both GDPR and ISO/IEC 29100. The requirements are therefore consistent, and merged as R41 COLLECT the personal data as necessary for specific purposes.

The following example is made up for the purpose of explanation to demonstrate the inconsistency between requirements. Assuming a statement states ``Any personal data can be freely collected without specifying a specific purpose for collection'', we have derived the requirement as ``COLLECT any personal data without a specific purpose''. This requirement would contradict with requirement R41 discussed above since the former does not require a specific purpose provided, while the latter does.

We merged in total 178 similar and duplicate requirements. We did not find any inconsistent requirements. For requirements traceability, we have provided a full list of the privacy requirements with their references to the GDPR articles, ISO/IEC 29100 principles, Thailand PDPA sections and APEC framework points in the replication package \cite{reppkg-pridp}. This step resulted in a final taxonomy of 71 privacy requirements in 7 goal categories which we will discuss in detail in the next subsection.

\newtext{\subsection{Privacy requirements classification}} \label{subsec:req-classification}

In this step, we aim to group the privacy requirements into categories based on their goals. \newtext{We adopted the empirical-to-conceptual approach to develop our taxonomy \cite{Nickerson}. We identified the privacy requirements that had common characteristics, grouped those privacy requirements together, and finally formed the categories. As we have classified the privacy requirements from their smallest unit of analysis (i.e. each privacy requirement), thus we called this approach as the bottom-up approach.} This approach ensures that the generated categories cover and address all the requirements. The approach also allows the categories in the taxonomy to be updated when there are new privacy requirements identified in the future. For example, newly identified privacy requirements can be added to existing categories or form new categories.

The bottom-up approach consists of two steps. We first considered the privacy requirements based on their actions, objects and target results. For example, the privacy requirements with the action verb \emph{ALLOW}, the object \emph{data subjects} and the target results that are related to individual rights (e.g. access, rectify and erase) were grouped together (i.e. user participation). Similarly, the privacy requirements with the action verb \emph{PROVIDE}, the object \emph{data subjects} and the target results that are related to information for stakeholders were gathered into the same group. We kept applying this strategy to group the rest of privacy requirements. We then ended up with fifteen categories in the first step.

Next, we grouped the privacy requirements that have at least two common components either the actions, objects or target results. For example, we gathered the privacy requirements that have the same action verb \emph{PROVIDE} and the same target results that are related to information for stakeholders, but there are two different objects - data subjects and other parties that are not data subjects. We then created subcategories for each object (i.e. notice - data subjects and notice - relevant parties) \newtext{(see Subcategories 2.1 and 2.2 in Table \ref{tab:sample-requirements})}. All of these requirements were grouped under the notice category. We repeated this step with the rest of privacy requirements. Finally, the taxonomy consists of seven categories (some of which have sub-categories): user participation, notice, user desirability, data processing, breach, complaint/request and security. 

After we had categorised all the privacy requirements, we noticed that some of them address more than one category. For example, requirement \emph{R6 ALLOW the data subjects to withdraw consent} addresses both user participation and also consent (which is under user desirability) categories. Thus, we again went through all the privacy requirements, and considered if the privacy requirements address other relevant categories other than their existing one. We added those privacy requirements into other relevant categories. Thus, some requirements can belong to multiple categories. The descriptions of privacy goal categories and the examples of privacy requirements in each goal category are explained in detail in Section \ref{sec:taxonomy}. \\

\newtext{We note that the mutual exclusivity may be desirable for some taxonomies, however it is not compulsory as long as the taxonomy is useful \cite{Nickerson}. We demonstrated our taxonomy is useful based on the following attributes:}

\begin{itemize}[leftmargin=*,noitemsep]
	\item \newtext{Concise: our taxonomy consists of limited categories to cover all privacy requirements.}
	\item \newtext{Robust: the categories in the taxonomy clearly and adequately differentiate the privacy requirements into specific and relevant groups.}
	\item \newtext{Comprehensive: our taxonomy can classify the privacy requirements within the concerned principles and properties of the data protection and privacy regulations and frameworks.}
	\item \newtext{Extensible: the approach we adopted in the study allows the categories in the taxonomy to be updated when there are new privacy requirements identified in the future.}
	\item \newtext{Explanatory: our categories have their own characteristics and are able to provide specific explanation to describe the privacy requirements under them.}
\end{itemize}

\begin{figure}[h]
	\centering
	\includegraphics[width=1\linewidth]{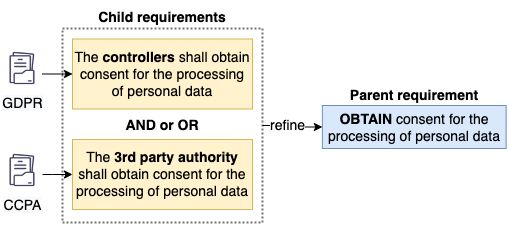}
	\caption{\newtext{An example demonstrates a scenario where two requirements aiming to achieve the same goal having different actors}}
	\label{fig:different-actors}
\end{figure}

Our methodology is also able to address the scenario where two requirements aiming to achieve the same goal have different actors. We can further refine relevant requirements into a parent requirement and child requirements with specific logical operators (i.e. AND and OR). \newtext{Since we did not find any case where two requirements aiming to achieve the same goal have different actors in the privacy requirements extracted from GDPR, ISO/IEC 29100 Thailand PDPA or APEC framework, we used a sample of privacy requirement extracted from other regulations to illustrate the privacy requirement related to obtaining consent instead.} For example, assume that the controller is required to obtain consent for the processing of personal data. The data controller is responsible for this task in GDPR, however this could be managed by 3rd party authority in other regulations (e.g. CCPA). We can then refine these requirements into a parent requirement as \textit{obtain consent for the processing of personal data}. The child requirements could be expressed as \textit{the controllers shall obtain consent for the processing of personal data} and \textit{the 3rd party authority shall obtain consent for the processing of personal data} (see Figure \ref{fig:different-actors}). The logical operator in this scenario is \textit{OR} as software development teams can choose between one of the child requirements to implement to satisfy the parent requirement.


If a regulation states that it requires both the controllers and the 3rd party authority to obtain consent for the processing of personal data, the logical operator in this scenario is \textit{AND}, and the software development teams must implement both child requirements in their system. As a data controller is the main actor in the regulations and frameworks we analysed in this study, the above scenario did not occur in our study. Thus, we did not refer to a specific requirement in our taxonomy.

\vspace{-4mm}

\section{\newtext{Privacy Requirements Taxonomy}} \label{sec:taxonomy}

This section presents a taxonomy of privacy requirements that we have developed based on the GDPR, ISO/IEC 29100, Thailand PDPA and APEC privacy framework. Our taxonomy consists of a comprehensive set of 71 privacy requirements classified into 7 categories. The full version of the taxonomy can be found in online Annex \cite{reppkg-pridp}. We now highlight some of the important requirements in each category (see Table \ref{tab:sample-requirements}). We note that there are typically four types of roles involved in a privacy requirement: (i) data subjects who provide their personal data for processing, give consent and determine their privacy preferences; (ii) data controllers who determine what data to be collected and the purpose of personal data collection and processing; (iii) data processors who process the personal data corresponding to the specified purpose and (iv) third parties who in case receive personal data from the controllers or processors.

\vspace{-4mm}

\begin{table}[htbp]
	\caption{Selected privacy requirements that are referred in the paper. The full taxonomy is available in the replication package \cite{reppkg-pridp}.}
	\label{tab:sample-requirements}
	\small
	\begin{tabular}{p{8.5cm}}
		\toprule 
		\textbf{Privacy requirements}\\
		\midrule 
		
		\textbf{Category 1: User participation} \\
		R1 ALLOW the data subjects to access and review their personal data \\
		R6 ALLOW the data subjects to withdraw consent \\
		R34 ALLOW the data subjects to obtain and reuse their personal data for their own purposes across different services \\
		R44 ALLOW the data subjects to erase their personal data     \\
		R45 ALLOW the data subjects to rectify their personal data \\
		
		\vspace{1mm}
		
		\textbf{Category 2: Notice} \\
		\newtext{\textbf{Subcategory 2.1: Data subjects}} \\
		R12 INFORM the data subjects the reason(s) for not taking action on their request and the possibility of lodging a complaint \\
		R15 NOTIFY the data subjects the data breach which is likely to result in high risk \\
		R19 PROVIDE the data subjects an option to choose whether or not to provide their personal data \\
		R22 PROVIDE the data subjects with the identity and contact details of a controller/controller's representative \\
		R26 PROVIDE the data subjects the information relating to the policies, procedures, practices and logic of the processing of personal data   \\
		R27 PROVIDE the data subjects the recipients/categories of recipients of their personal data   \\
		R30 PROVIDE the data subjects the information relating to the processing of personal data with standardised icons \\
		R38 PROVIDE the data subjects the purpose(s) of the collection of personal data      \\
		R39 PROVIDE the data subjects the purpose(s) of the processing of personal data \\
		R42 PROVIDE the data subjects the categories of personal data concerned  \\			
		R55 PROVIDE the data subjects the period/criteria used to store their data \\
		
		\newtext{\textbf{Subcategory 2.2: Relevant parties}} \\
		R17 SHOW the relevant stakeholders the consent given by the data subjects to process their personal data  \\
		R66 NOTIFY a supervisory authority the data breach   \\
		
		\vspace{1mm}
		
		\textbf{Category 3: User desirability} \\
		\newtext{\textbf{Subcategory 3.1: Consent}} \\
		R6 ALLOW the data subjects to withdraw consent \\
		R8 IMPLEMENT the data subject's preferences as expressed in his/her consent \\
		R35 OBTAIN the opt-in consent for the processing of personal data for specific purposes \\
		R47 ERASE the personal data when a consent is withdrawn \\
		
		\newtext{\textbf{Subcategory 3.2: Choice}} \\
		R19 PROVIDE the data subjects an option to choose whether or not to provide their personal data  \\
		R36 PRESENT the data subjects an option whether or not to allow the processing of personal data \\
		
		\newtext{\textbf{Subcategory 3.3: Preference}} \\
		R8 IMPLEMENT the data subject's preferences as expressed in his/her consent \\
		
		\vspace{1mm}
		
		\textbf{Category 4: Data processing} \\	
		\newtext{\textbf{Subcategory 4.1: Collection}} \\		
		R41 COLLECT the personal data as necessary for specific purposes   \\

		\bottomrule
		
	\end{tabular}
\end{table}

\begin{table}[h]
	\label{tab:sample-requirements-2}
	\small
	\begin{tabular}{p{8.5cm}}
		\toprule 
		\textbf{Privacy requirements (Continued)}\\
		\midrule 
		
		
		\newtext{\textbf{Subcategory 4.2: Use}} \\		
		R40 USE the personal data as necessary for specific purposes specified by the controller \\
		
		\newtext{\textbf{Subcategory 4.3: Storage}} \\		
		R43 STORE the personal data as necessary for specific purposes \\
		
		\newtext{\textbf{Subcategory 4.4: Erasure}} \\		
		R7 ERASE the personal data when it has been unlawfully processed \\
		R46 ERASE the personal data when the data subjects object to the processing \\
		R47 ERASE the personal data when a consent is withdrawn \\
		R52 ERASE the personal data when it is no longer necessary for the specified purpose(s)   \\
		R53 ERASE the personal data when the purpose for the processing has expired  \\
		
		\newtext{\textbf{Subcategory 4.6: Record}} \\		
		R13 MAINTAIN a record of personal data processing activities \\
		
		\vspace{1mm}
		
		\textbf{Category 5: Breach} \\
		R15 NOTIFY the data subjects the data breach which is likely to result in high risk \\
		R66 NOTIFY a supervisory authority the data breach   \\
		R67 NOTIFY relevant privacy stakeholders about a data breach \\
		
		\vspace{1mm}
		
		\textbf{Category 6: Complaint/Request} \\
		R12 INFORM the data subjects the reason(s) for not taking action on their request and the possibility of lodging a complaint \\
		R31 REQUEST the data subjects the additional information necessary to confirm their identity when making a request relating to the processing of personal data \\
		
		\vspace{1mm}
		
		\textbf{Category 7: Security} \\
		R56 ALLOW the authorised stakeholders to access personal data as instructed by a controller \\
		R60 IMPLEMENT appropriate technical and organisational measures to protect personal data \\
		R63 PROTECT the personal data from unauthorised access and processing \\
		R65 IMPLEMENT a function to comply with local requirements and cross-border transfers \\
		\bottomrule
	\end{tabular}
\end{table}


\newtext{\subsection{User participation}}
\newtext{All the requirements in this category specify the functionalities provided for data subjects to execute their individual rights in managing their personal data. The data subjects must be able to access and review, erase and rectify their personal data (e.g. R1, R44 and R45). The systems must allow the data subjects to withdraw consent (R6). The systems must also provide the data subjects their personal data when they would like to obtain and reuse their personal data for their own purposes across different services (R34). The controllers shall allow the data subjects to object to and restrict the processing of their personal data (R3 and R4). The data subjects must be able to withdraw consent (R6) or lodge a complaint to a supervisory authority (R2).}

\newtext{\subsection{Notice}}
\newtext{This category is the largest group consisting of 32 privacy requirements in the taxonomy. It consists of two sub-categories: data subjects and relevant parties. Most of the requirements in this category are concerned with the transparency of personal data processing (e.g. R17, R22 and R42). It has a set of requirements for the data subjects to be informed and/or notified of relevant information and individual rights related to the processing of personal data. Those requirements aim to ensure that a system shall provide information related to the processing of personal data (e.g. what personal data is required, who is responsible for their data and results from requests) to data subjects. The information also includes privacy policies, procedures, practices and logic of the processing of personal data. Personal data shall not be misused, and the data subjects have the right to know the purpose of collection and processing (R38 and R39). The data subjects should be provided the duration their personal data will be stored (R55). Additional information must be provided to the data subjects if the collected personal data are required for other purposes (R37). General privacy-related information should be presented in a clear and simple, accessible language without technical terms as required in requirement (R26). Any updates of personal data processing must be informed to the recipients (i.e. processors or third parties) of those personal data (R50). The controllers must provide the contact details of responsible persons who control the processing (R20 and R22). The data subjects must be notified when they are likely to be in risk from personal data exposures (R15).}

\newtext{\subsection{User desirability}}
\newtext{This category consists of 9 requirements categorised into three sub-categories: consent, choice and preferences. The requirements in this category focus on ensuring that the processing of personal data is performed according to data subjects' consent and preferences. A number of requirements focus on the controllers being given authorities to process personal data (e.g. R8). It is also necessary to obtain consent for the processing based on those purposes (R35). The data subjects have options to allow the processing of their personal data for a certain specific purpose (R36).}


\newtext{\subsection{Data processing}}

\newtext{This category addresses the processing of personal data from the controllers' side (16 requirements). The sub-categories in this category include collection, use, storage, erasure, transfer and record. The controllers are expected to collect and store only personal data that is required in the processing for the specific purpose(s) (R41 and R43). A set of requirements involves data erasure in systems. Requirement R53 addresses the case of removing personal data when the purpose for processing has expired. When the data subjects would like to have their personal data erased, the system shall provide this processing lawfully (R46 and R47). When the processing is complete and the personal data is no longer needed, the personal data should be removed from the system unless they are required by law/regulations (e.g. R51 and R52). In case that the personal data is unlawfully processed, the data must be removed from the systems (R7). In addition, personal data must be used only for the specified purpose(s) (R40). When requested by the data subjects, the controllers must transmit their personal data to another controller (R33). The data subject must be informed when their personal data needs to be transferred to a third country or an international organisation (R9). The controllers shall document the categories of personal data collected as it is important to know what personal data are stored in the systems (R70).}

\newtext{\subsection{Breach}}
This goal category focuses on providing and notifying important information related to personal data breaches to data subjects, relevant stakeholders and a supervisory authority (e.g. R15 and R71). Thus, it is important to implement a functionality that satisfies this compliance in the systems (e.g. R66). Requirement R67 imposes good practices of informing the related parties about the breaches. The controllers must be informed by processors about the breaches as well (R68). The controllers shall document the details of data breaches for verifying their compliance (R69).

\newtext{\subsection{Complaint/Request}}
\newtext{This privacy goal consists of 5 privacy requirements. It concerns complaint and request made by both data subjects and controllers. If the controllers refuse to take actions on the data subjects' requests about their individual rights, they have to provide a reason to the data subjects (R12). The data subjects must be able to lodge their complaints with a supervisory authority (R2). The controllers shall process personal data as requested (e.g. transmit personal data to another controller) (R33). The controllers should request additional information to confirm data subjects' identity when requests have been made (R31).}

\newtext{\subsection{Security}}
There are 13 requirements in our taxonomy covering the security practices in maintaining integrity, confidentiality and availability. The systems must allow only authorised people to access or process personal data (R56). The personal data should be protected with proper mechanisms (e.g. R60 and R63). The systems must restore the availability and access to personal data after incidents (R62). The interactions in the systems should neither identify nor observe the behaviour of the data subjects as well as reduce the linkability of the personal data collected (R64). Apart from the fundamental practices that the personal data should be protected, this is beneficial when the personal data is exposed. The data protection approaches such as anonymisation and pseudonymisation can help reduce the impact of privacy breaches. Most importantly, a set of requirements require the systems to implement mechanisms to ensure security and privacy compliance (e.g. R57, R58, R61 and R66). In addition, the implementation of mechanisms to assess the accuracy and quality of procedures should be considered (R49). 

In case that the personal data are processed across organisations/countries, the controllers must ensure local requirements and cross-border transfers (R65). Cross-border transfers are challenging for both controllers and processors. In cross-border transfer settings, it is required that the requirements at the destination should be equivalent to the ones at the source. The processors outside EU are sometimes not aware of those scenarios since they may not normally process personal data of EU citizens and residents. Therefore, the controllers are responsible for verifying requirements compliance before transferring personal data.

\newtext{All categories mentioned above can be considered as components in a software system. The privacy requirements are tasks in each component that the software engineers and relevant stakeholders should consider when designing and developing a system. Implementing those privacy requirements will fulfill the needs of stakeholders, which in this case is to satisfy the requirements stated in privacy and data protection regulations and frameworks.} \\

\section{\newtext{Identifying privacy requirements in issue reports}} \label{sec:mining}

Most of today's software projects follow an agile style in which the software development is driven by resolving issues in the backlog. In those projects, issue reports contain information about requirements of a software that are recorded in multiple forms in an issue tracking system \cite{Choetkiertikul}: new requirements (such as user stories or new feature request issues), modification of existing requirements (such as improvement request issues) or reporting requirements not being properly met (i.e. bug report issues). Large projects may have thousands of issues which provide fairly comprehensive source of requirements about the projects and the associated software systems. Thus, we have performed a study on the issue reports of software projects to understand how the selected projects address the privacy requirements in our taxonomy. As issue reports can be in multiple forms, we note that the taxonomy can be applied to software requirements and user stories. In this section, we first describe the process steps that we have followed to carry out our study, assess classification results, resolve disagreements, and discuss the outcomes.

\begin{figure}[ht]
	\centering
	\includegraphics[width=.9\linewidth]{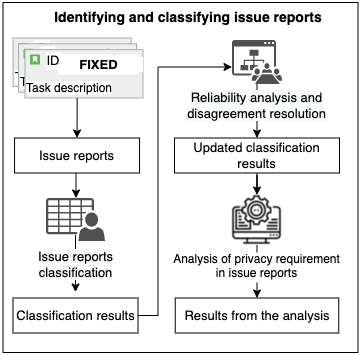}
	\caption{An overview process of identifying and classifying privacy requirements in issue reports}
	\label{fig:Mining issue reports}
\end{figure}

\subsection{Issue reports collection}

We apply a number of criteria to select software projects for this study as follows: (i) are open-source projects, (ii) serve a large number of users, (iii) are related to privacy and (iv) have accessible issue tracking systems. There were a number of projects which satisfy these criteria. Among them, we selected Chrome and Moodle due to their large scale size, popularity and representativeness\footnote{As discussed later in the paper, performing thorough analysis as we have done on these two projects alone required significant effort (278 person-hours). Hence, we scoped our study in these two projects so that we can publish this dataset timely, enabling the community to initiate research on this important topic, and subsequently extend it with additional projects and data.}. Google Chrome is one of the most widely-used web browsers, which was developed under the Chromium Projects \cite{Projects}. As a web browser, Chrome stores personal data of users, e.g. username, password, address, credit card information, searching behaviour, history of visited sites and user location. Moodle is a well-known open-source learning platform \cite{Moodle} with over 100 million users worldwide. Moodle aims to comply with GDPR \cite{Moodle2019}.

Figure \ref{fig:Mining issue reports} shows the process steps that we have followed in our study. We first identified privacy-related issues from all the issue reports we collected from Chrome and Moodle projects. To do so, we identified issues that were explicitly tagged as privacy in the ``component'' field of both projects and their status was either assigned, fixed and verified (to ensure that they are valid issues). This is to ensure that the issue reports we collected were explicitly tagged as ``privacy'' by Chrome and Moodle contributors. This process initially gave us 1,080 privacy-related issues from Chrome and 524 from Moodle. We then manually examined those issues and filtered out those that have limited information (e.g. the description that does not explain the issue in detail or contain only source code) to enable us to perform the classification. For example, issue ID 953622\footnote{https://bugs.chromium.org/p/chromium/issues/detail?id=953622} states that ``Null-dereference READ in bool base::ContainsKey<std::\_\_Cr::map<std::\_\_Cr::basic\_string<char, std::\_\_Cr::c''. This issue description is very brief and does not contain any explanation describing what the issue is about, what personal data is concerned or which function is affected in the issue. We thus exclude the issue from our study. Finally, our data contains 896 issues from Chrome and 478 issues from Moodle.

In Chrome dataset, the collected issue reports were created between January 2009 and March 2020. There are five issue types reported in our dataset: bug, bug-regression, bug-security, feature and task. Each issue report has seven contributors on average; the contributors include reporters, owners and relevant collaborators. Issue reports in Moodle dataset span for two years, 2018 - 2019. The issue types in Moodle include bug, epic, improvement, new feature, task, sub-task and functional test. On average, five participants are involved in each report including reporters, assignees, testers and commenters. The descriptive statistics of the issue reports can be seen in Table \ref{tab:issue-stats}\footnote{The number of issue reports counted by issue type in both projects is provided in the supplementary material.}.

\begin{table*}[h]
	\centering
	\caption{Descriptive statistics of the number of contributors, resolution time and number of comments of the issue reports in our datasets.}
	\label{tab:issue-stats}
	\resizebox{6.5in}{!}{%
		\begin{tabular}{@{}llllllllllllllll@{}}
			\toprule
			\multirow{2}{*}{\textbf{Project}} & \multicolumn{5}{c}{\textbf{\#Contributors}} & \multicolumn{5}{c}{\textbf{Resolution Time (days)}} & \multicolumn{5}{c}{\textbf{\#Comments}} \\ \cmidrule(l){2-16}
			& \textbf{min} & \textbf{max} & \textbf{mean} & \textbf{median} & \textbf{mode} & \textbf{min} & \textbf{max}  & \textbf{mean} & \textbf{median} & \textbf{mode} & \textbf{min} & \textbf{max} & \textbf{mean} & \textbf{median} & \textbf{mode} \\ \midrule
			Google Chrome & 1   & 32  & 5    & 4      & 2    & 1   & 3,635 & 315  & 65     & 1    & 0   & 311 & 16   & 12     & 12    \\
			Moodle        & 1   & 14  & 4    & 5      & 5    & 1   & 852   & 37   & 13     & 1    & 0   & 112 & 11   & 9     & 1    \\ \bottomrule
		\end{tabular}%
	}
	{\parbox{16.5cm}{\footnotesize \#Contributors: number of contributors, \#Comments: number of comments, min: minimum of contributors/resolution time/comments, max: maximum of contributors/resolution time/comments, mean: mean of contributors/resolution time/comments, median: median of contributors/resolution time/comments, mode: mode of contributors/resolution time/comments.}}
\end{table*}

\subsection{Issue reports classification}

In this phase, we went through each issue report in the dataset to classify it into the privacy requirements in our taxonomy. This phase consists of three steps: (i) identifying concerned personal data described in the issue report, (ii) identifying functions/properties reported in the issue, (iii) mapping the issue to one or more privacy requirements. 

Regarding the classification, each coder was initially provided with an online form containing the title and description of the assigned issue reports and 71 columns representing each requirement. The coders analysed each issue following the classification steps described above (i.e. steps (i) - (iii)). The coders carefully consider every scenario mentioned in the issue reports. Once the coders have identified related information about personal data and function(s) concerned, they considered the relevant requirements. The coders determined the requirement(s) that matches with information analysed above. The coders then updated the value in the columns of chosen requirement(s) in the given form. Finally, all the coders delivered their result file containing the issue reports and their privacy requirement labels for reliability assessment process. To ensure that the classification process is reliable, two coders were assigned to classify an issue report. The reliability assessment and disagreement resolution processes are described in detail in Section \ref{subsec:mining-reliability}.

The following example demonstrates the issue reports classification in our datasets. Issue 123403\footnote{https://bugs.chromium.org/p/chromium/issues/detail?id=123403} in Chrome reports that \textit{``Regression: Can't delete individual cookies''}. The personal data affected here is individual cookies, and the function reported is erasing or deleting (individual cookies). Thus, we classify this issue into the requirement \textit{\textbf{ALLOW} the data subjects to erase their personal data (R44)} in our taxonomy (see Table \ref{tab:sample-requirements}). The users should be able to select the cookies that they want to delete.

In another example, issue 495226\footnote{https://bugs.chromium.org/p/chromium/issues/detail?id=495226} in Chrome requests that the \textit{``Change Sign-in confirmation screen''} should be changed. The description of this issue requires that the system should inform the reasons for user account data collection and how this data will be further processed before obtaining this data in the sign-in process. Since this issue requires that the user should be informed of the purpose of collection and processing, the issue can be classified into requirements R38 and R39 (see Table \ref{tab:sample-requirements}). Both of these requirements belong to the notice privacy goal. This example shows that an issue can be classified into more than one privacy requirements.

Issue 831572\footnote{https://bugs.chromium.org/p/chromium/issues/detail?id=831572} in Chrome requires: \textit{``Provide adequate disclosure for (potentially intrusive) policy configuration''}. Further investigation into the issue's description revealed that the disclosure of policy configuration includes: letting the users know that they are managed, and providing indication when user data may be intercepted and when user actions are logged locally. These involve the following functions: (i) the users should be informed of the purpose of processing so that they know they are managed; (ii) the enterprise may intercept the users' data, thus the users should know whom their personal data might be sent to; and (iii) the history of user logging shall be recorded to acquire logging data. Hence, this can be classified into three requirements R39, R27 and R13 (see Table \ref{tab:sample-requirements}). This example demonstrates that one issue relates to several requirements across different privacy goal categories: notice and data processing.

The following example demonstrates the issue that concerns multiple functions which more than one privacy goal is addressed in Moodle. Issue MDL-62904\footnote{https://tracker.moodle.org/browse/MDL-62904} in Moodle reports that \textit{``users can't find where to request account deletion''}. The issue was described that the system does not provide a function for users to request for deleting their account in the user interface. Hence, this issue addresses requirements R30 in the notice category and R44 in the user participation category in the taxonomy.

Although the mapping is relatively straightforward in most of the issues, some presents challenges. For instance, a set of issues in Moodle refer to the implementation of the core\_privacy plugins (e.g. MDL-61877\footnote{https://tracker.moodle.org/browse/MDL-61877}). However, the information given in the description of those issues is inadequate to identify which privacy requirements are related to. Therefore, we needed to seek for additional information about core\_privacy plugins in Moodle development documentation \cite{Moodle2019}. In addition, the issues in Chrome and Moodle projects have specific function names or technical terms used by their developers. Hence, extra effort was required to understand those issues and classify them. Once the coders acquired the additional information from software documentation, they discussed potential privacy issues/functionalities raised in those issue reports. For example, the documentation of core\_privacy plugins explains six functionalities that the plugins should provide. The coders then discussed which privacy requirements in the taxonomy involved with those functionalities, and classified that issue according to the identified privacy requirements.

\subsection{Reliability analysis} \label{subsec:mining-reliability}

The classification process has been performed by three coders, who also involved in the taxonomy development process. All the coders have independently followed the above process to classify issue reports into our taxonomy of privacy requirements. The first coder was responsible for classifying all 1,374 privacy issue reports. The classification process is however labour intensive. It took the first coder approximately 138 person-hours to classify all 1,374 issue reports (6 minutes per issue report in average). The second and third coders spent approximately 70 hours per person to classify 687 issue reports. In total, the process took four months to complete, so the coders had time to manage their workloads efficiently. The coders usually divided the assigned issue reports into smaller sets (i.e. about 50 issue reports) to work on for each round. For the purpose of reliability assessment, the second and third coder each was assigned to classify a different half of the issues in each project. This setting aimed to ensure that each issue report is classified by at least two coders.

An issue report can be classified into multiple privacy requirements (i.e. a multi-labelling problem). Hence, we employ Krippendorff's alpha coefficient \cite{Krippendorff2011}, \cite{Artstein2008} with MASI (Measurement Agreement on Set-valued Items) distance to measure agreement between coders with multi-label annotations \cite{Ravenscroft2016}, \cite{Passonneau2006}. The MASI distance measures difference between the sets of labels (i.e. privacy requirements) provided by two coders for a given issue. The Krippendorff's alpha values between the first and second coders are 0.509 for Chrome and 0.448 for Moodle. The agreement values between the first and third coder are 0.482 and 0.468 for Chrome and Moodle respectively. A disagreement resolution step was conducted to resolve the classification deviations between the three coders\footnote{Krippendorff’s alpha values indicate the degree of (dis-)agreement between coders. In our case, disagreements were often due to the ambiguity in the issue reports, requiring us to perform a resolution step. Doing these is to increase the reliability of our dataset.}.

The low Krippendorff's alpha values from the initial classification were \emph{not} due to the privacy requirements in our taxonomy. Rather, it was mainly because of the limited information provided in the description of a number of issue reports, forcing the coders to make their own assumptions about the nature of those issues. If an issue report is clearly described, the coders classified it into the same requirements. 53.01\% of the Chrome issues and 46.23\% in Moodle received this total agreement between all the three coders. We have addressed this problem by conducting a disagreement resolution step where all the coders met and discussed to resolve the disagreements. 


\newtext{\textbf{Disagreement resolution}:} \newtext{We conducted several meeting sessions between the coders to resolve disagreements in a sample dataset. The sample dataset contained the issue reports that were classified into different privacy requirements by both coders. Specifically, both coders did not classify those issue reports into at least one same requirement. There were 343 and 161 issue reports in Google Chrome and Moodle resolved in the meeting sessions between the coders who were responsible for the classification.} Several meetings were conducted because of the time difference and availability among the coders. Each meeting took 3 hours on average as we revisited, discussed and reclassified the issues with disagreements issue by issue. The same coders continue working on the disagreed issues in the same set of issues they had annotated in the previous step. During these meetings, the coders examined the issues thoroughly (not just only their description, but also other documentation related to the issues), discussed to develop a mutual understanding of the issue, and then reclassified the issue together. For each issue in the list, each coder explained the justification for their classification of the issue. The coders then resolved the disagreements on that issue in two ways. After the discussion, if the coders agreed with the other coder's classification, the labels were combined (i.e. the issue were classified into multiple requirements). If the coders did not reach an agreement, they went through the issue's description together to discuss and develop a mutual understanding of the issue. They then reclassified the issue together. Hence, the final classification has maximum agreement among the coders, thus ensuring the reliability of our dataset. \newtext{Once all disagreements had been resolved for the sample set, the first coder adjusted the classification of the issues which were not included in the sample set to finalise the datasets. This included 78 Chrome and 96 Moodle issue reports respectively.}

The following example illustrates how conflicts between coders were resolved. Issue 527346\footnote{https://bugs.chromium.org/p/chromium/issues/detail?id=527346} in Chrome requests that the users should know when they are managed. The description of the issue requires the system to show information to users when they are managed and the information should be easily seen by users. This issue was classified to R26 by one coder and R30 by the other coder. Although both R26 and R30 involve providing information to users, R26 focuses on the information relating to the policies, procedures, practices and logic of the processing of personal data, while R30 focuses on representing information relating to the processing of personal data with standardised icons in the user interface. After the coders revisited the issue and discussed the description in detail, the coders agreed that the information should be shown in the tray bubble which is a part of the user interface. The coders therefore reclassified this issue into R30.

\section{Analysis and discussions} \label{sec:results}

The study presented in the previous section generates not only a valuable dataset but also important insights into how privacy requirements have been addressed in Chrome and Moodle issue reports. In this section, we discuss and analyse some of the key findings and implications.


\subsection{The top and least concerned privacy requirements}

Table \ref{tab:req-occurrences} presents the coverage of each category in Chrome and Moodle issue reports. \newtext{We found 1,157 and 1,275 privacy requirements in Chrome and Moodle issue reports respectively (2,432 privacy requirements in total).} In both projects, the majority of the issues address the user participation requirements (Category 1). Most issue reports in Moodle address more than one privacy requirement across different privacy goal categories. This results in Moodle having higher coverage (in terms of the occurrences) than Chrome. \newtext{It is worth noting that requirements R1 ALLOW the data subjects to access and review their personal data, R26 PROVIDE the data subjects the information relating to the policies, procedures, practices and logic of the processing of personal data, R30 PROVIDE the data subjects the information relating to the processing of personal data with standardised icons, R44 ALLOW the data subjects to erase personal data and R60 IMPLEMENT appropriate technical and organisational measures to protect personal data were in the top 10 in both projects.}

\begin{table}[ht]
	\centering
	\caption{\newtext{The number of privacy requirements found in Chrome and Moodle issue reports categorised by category.}}
	\label{tab:req-occurrences}
	\resizebox{3.5in}{!}{%
		\begin{tabular}{@{}l c c@{}}
			\toprule
			\multicolumn{1}{c}{\multirow{2}{*}{\textbf{Category}}} & \multicolumn{2}{c}{\textbf{\begin{tabular}[c]{@{}c@{}}No. of mined privacy requirements \\ in each project by category\end{tabular}}} \\ \cmidrule(l){2-3}
			\multicolumn{1}{c}{}                                   & \multicolumn{1}{c}{\textbf{Google Chrome}}      & \multicolumn{1}{c}{\textbf{Moodle}}      \\ \midrule
			1) User participation							&   321                                              &   328                                       \\
			2) Notice                                              &   272                                              &   229                                       \\
			3) User desirability							&    241                                             &    194                                      \\
			4) Data processing 							&     117                                            &  55                                        \\
			5) Breach 											&    0                                             &     0                                     \\
			6) Complaint/Request 						&   1                                              &   9                                       \\
			7) Security 										&    160                                             &   214                                       \\ \bottomrule
		\end{tabular}%
	}
\end{table}

\begin{figure}[ht]
	\centering
	\includegraphics[width=1\linewidth]{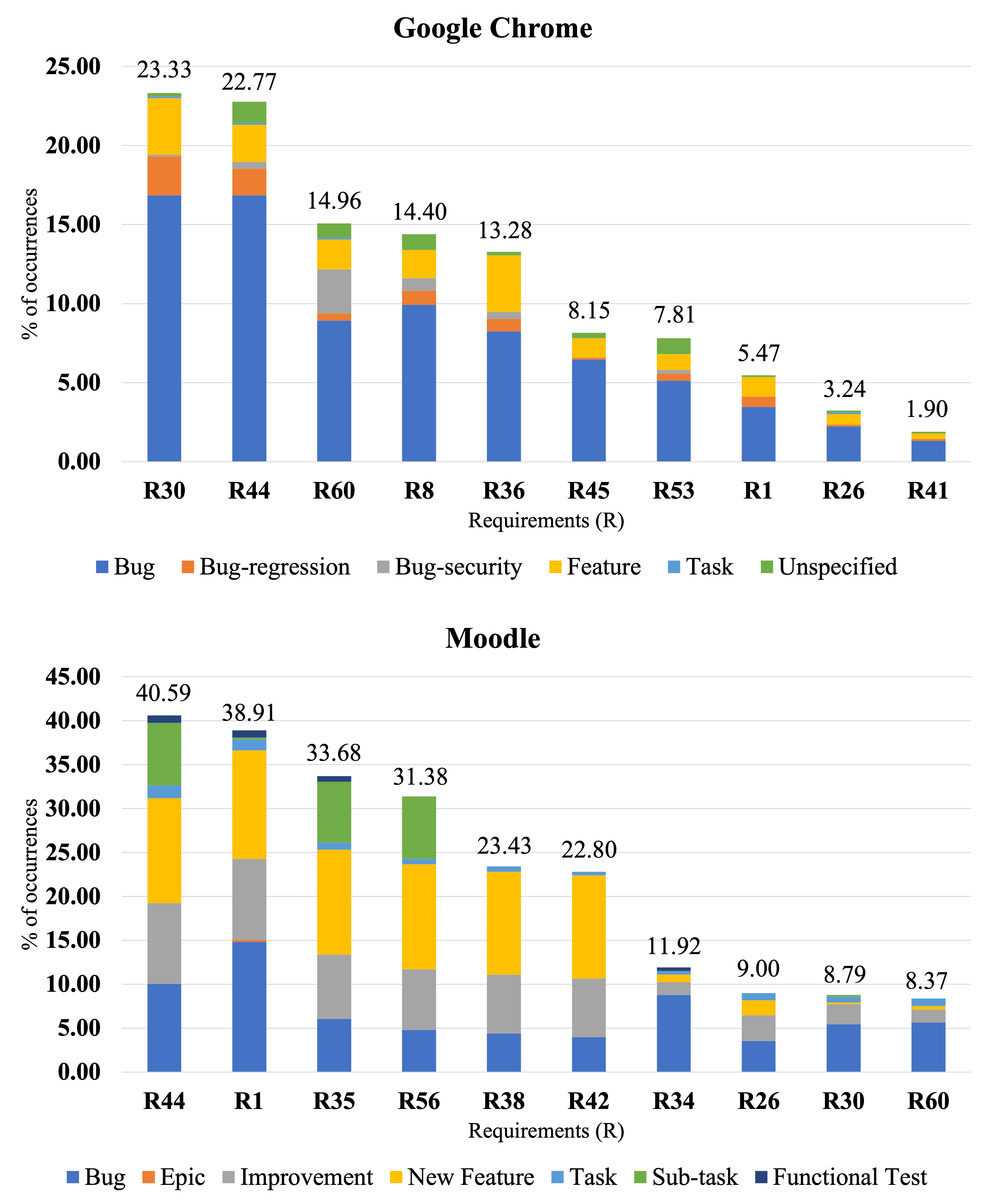}
	\caption{Top 10 privacy requirements occurrences in Google Chrome and Moodle datasets categorised by issue types}
	\label{fig:top10}
\end{figure}

The top three most concerned requirements in Chrome are R30, R44 and R60 (see Figure \ref{fig:top10} and \newtext{Table \ref{tab:top10-detail}}). Note that these requirements belong to three different privacy goal categories (refer to Table \ref{tab:sample-requirements} for details of the privacy goals and requirements we discussed here). The top three requirements covered in Moodle issues are R44, R1 and R35 \newtext{OBTAIN the opt-in consent for the processing of personal data for specific purposes}. 

Requirement R44 was in the top three most concerned requirements in both projects, suggesting that allowing the data subjects to erase their personal data is a highly important privacy requirement for both Chrome and Moodle. Requirements R30 and R36 were also addressed in many privacy-related issues in Chrome. This suggests that providing information with standardised, visible and meaningful icons which inform the intended processing of personal data for users is an important privacy concern in Chrome (R30). In addition, many issues in Chrome also focus on addressing the privacy requirement that users are presented with all available options related to the processing of personal data (R36).

Apart from requirement R44, the other two requirements most frequently covered in Moodle issue reports are R1 and R35 (note that they are different from those in Chrome). Approximately 39\% issues in Moodle are related to requirement R1. This implies that Moodle has a strong emphasis on allowing users to access their personal data such as grade records, course participation and course enrolment records. This function is not only important for Moodle, but also in Chrome (R1 is also in the top 10 for Chrome). User consent is a major concern in privacy protection. We found that a large number of issue reports in Moodle explicitly requires the system to obtain consent from users for processing personal data based on specific purposes (R35).


It is interesting to note that none of the requirements in the breach category were found in both Chrome and Moodle as they were not directly observed from issue reports or recorded in the ITS. These goals can be evidenced through high-level organisational activities such as Data Protection Impact Assessment, Legitimate Interest Assessment and breach notifications. Our future work will investigate this further.

\begin{table}[ht]
	\centering
	\caption{\newtext{A summary of top 10 concerned privacy requirements in Google Chrome and Moodle datasets.}}
	\label{tab:top10-detail}
	\resizebox{3.5in}{!}{%
	\begin{tabular}{@{}ccllc@{}}
		\toprule
		\textbf{Project}                                      & \textbf{\begin{tabular}[c]{@{}c@{}}Req-\\ uirement\end{tabular}} & \multicolumn{1}{c}{\textbf{Category}} & \multicolumn{1}{c}{\textbf{Subcategory}}                               & \textbf{\begin{tabular}[c]{@{}c@{}}Frequency\\ (Occurrences)\end{tabular}} \\ \midrule
		\multirow{10}{*}{\textbf{Chrome}}                     & R30                  & 2) Notice                             & 2.1) Data subjects                                                     & 209                                                                        \\
		& R44                  & 1) User participation                 & -                                                                      & 204                                                                        \\
		& R60                  & 7) Security                           & -                                                                      & 135                                                                        \\
		& R8                   & 3) User desirability                  & \begin{tabular}[c]{@{}l@{}}3.1) Consent\\ 3.3) Preference\end{tabular} & 144                                                                       \\
		& R36                  & 3) User desirability                  & 3.2) Choice                                                            & 119                                                                        \\
		& R45                  & 1) User participation                 & -                                                                      & 73                                                                         \\
		& R53                  & 4) Data processing                    & 4.4) Erasure                                                           & 70                                                                         \\
		& R1                   & 1) User participation                 & -                                                                      & 49                                                                         \\
		& R26                  & 2) Notice                             & 2.1) Data subjects                                                     & 29                                                                         \\
		& R41                  & 4) Data processing                    & 4.1) Collection                                                        & 17                                                                         \\ \midrule
		\multicolumn{1}{l}{\multirow{10}{*}{\textbf{Moodle}}} & R44                  & 1) User participation                 & -                                                                      & 194                                                                        \\
		\multicolumn{1}{l}{}                                  & R1                   & 1) User participation                 & -                                                                      & 186                                                                        \\
		\multicolumn{1}{l}{}                                  & R35                  & 3) User desirability                  & 3.1) Consent                                                           & 161                                                                        \\
		\multicolumn{1}{l}{}                                  & R56                  & 7) Security                           &  -                                                                      & 150                                                                        \\
		\multicolumn{1}{l}{}                                  & R38                  & 2) Notice                             & 2.1) Data subjects                                                     & 112                                                                        \\
		\multicolumn{1}{l}{}                                  & R42                  & 2) Notice                             & 2.1) Data subjects                                                     & 109                                                                        \\
		\multicolumn{1}{l}{}                                  & R34                  & 1) User participation                 & -                                                                      & 57                                                                         \\
		\multicolumn{1}{l}{}                                  & R26                  & 2) Notice                             & 2.1) Data subjects                                                     & 43                                                                         \\
		\multicolumn{1}{l}{}                                  & R30                  & 2) Notice                             & 2.1) Data subjects                                                     & 42                                                                         \\
		\multicolumn{1}{l}{}                                  & R60                  & 7) Security                           & -                                                                      & 40                                                                         \\ \bottomrule
	\end{tabular}%
}
\end{table}

We have performed further analysis on the issue reports that were classified into the top concerned requirements in Chrome and Moodle datasets. We analyse four factors focusing on how the contributors treat those issue reports: issue types, the time took to resolve an issue, the number of contributors involved and the number of comments associated with the issue.

In Chrome dataset, issue reports have five different types: bug, bug-regression, bug-security, feature and task. Bug type reports malfunctioning functionalities in current version of the system. Bug-regression focuses on the functions that used to work correctly in the previous versions, but are broken in the current version. Bug-security reports malfunctions that are risky to user security. Feature type requests for an implementation of a new function/feature. Task type, which is not a bug or feature, defines a piece of work that needs to be completed for an issue. A small group of issues did not have their issue types specified. From our study, we found that the issue reports whose issue type is a bug are the largest group in every top concerned privacy requirement \newtext{(see Table \ref{tab:issue-type})}. These bug issue reports as well as bug-regression and bug-security took less time to resolve comparing to feature request issue reports on average for all the top concerned privacy requirements. In addition, the bug issue reports usually involved with a smaller number of contributors and had less discussions than the feature request issue reports. We have also investigated the discussions of bug issue reports classified into the top concerned requirements. We found that the bug issues were fixed after fifteen comments. However, the discussions of feature issues contain more details (e.g. use case scenarios, discussion points, screenshots and code snippets) than the bug issues.

There are seven different issue types in Moodle dataset: bug, epic, improvement, new feature, task, sub-task and functional test. The definitions of bug, new feature, task and sub-task issue types are similar to those mentioned in Chrome dataset. Epic issue type collects a group of issues that needs to be completed over a period of time. An improvement issue type is an enhancement to an existing feature. Functional test type contains the information and steps used for testing a particular function. From the analysis in Moodle dataset, the bug issue type contains the largest number of issue reports, followed by the feature issue type. We found that the bug issues did not only report the malfunctions, but they also reported the missing functionalities in the system (e.g. implement core\_privacy for block rss client). It is interesting to note that the new feature issues took less time to resolve comparing to the bug issues on average for all the requirements except R42. However, both issue types have similar number of comments (i.e. 11 to 12 comments) and number of contributors (i.e. 5).

\begin{table}[h]
	\centering
	\caption{\newtext{Number of issue reports in Chrome and Moodle projects counted by issue type.}}
	\label{tab:issue-type}
	\resizebox{3.5in}{!}{
	\begin{tabular}{p{2.5cm} p{1cm} p{2.5cm} p{1cm}}
		\toprule
		\textbf{Chrome} & \textbf{\#issues} & \textbf{Moodle} & \textbf{\#issues} \\
		\midrule
		Bug & 620 & Bug & 223 \\
		Bug-regression & 59 & Epic & 3 \\
		Bug-security & 36 & Improvement & 101 \\
		Feature & 132 & New Feature & 75 \\
		Task & 5 & Task & 26 \\
		Unspecified & 44 & Sub-task & 37 \\
		~ & ~ & Functional test & 13 \\
		\bottomrule
		\textbf{Total} & \textbf{896} & \textbf{Total} & \textbf{478} \\
		\bottomrule
	\end{tabular}%
	}
\end{table}

\newtext{We have summarised and presented the number of mined privacy requirements by category and by top requirements in both Chrome and Moodle projects\footnote{A full analysis can be found in the replication package \cite{reppkg-pridp}.}. These categories represent a group of activities/functionalities related to privacy in the projects while the privacy requirements identify specific needs expected to be fulfilled in a system. From our analysis, it presented how frequent those activities were concerned in those projects. This result allows the contributors to carefully consider the privacy-related functionalities in their projects. Software engineers can easily use the privacy requirements in the taxonomy to identify specific functionalities that were malfunctioned (e.g. bugs), needed to be implemented (e.g. feature requests) or needed to be changed (e.g. change requests) in issue reports. They can be confident that those functionalities are required by the data protection and privacy regulations, standards and frameworks. The privacy requirements mapped to each issue report can be also used as privacy measures. The contributors can assess whether the privacy measures are passed or failed. If the privacy requirements are not implemented or not properly functioned, then this issue report is failed for this project. On the other hand, if the privacy requirements are implemented or properly worked, then this issue report is passed. However, we did the analysis with the fixed issue reports, thus we could not assess privacy measures in those projects.}

\newtext{The top 10 privacy requirements in Chrome project covered 5 categories which are user participation, user desirability, notice, security and data processing (sorted by frequency). This set of privacy requirements covered only the data subjects subcategory in the notice category and collection and erasure subcategories in the data processing category. However, all the subcategories in user desirability category were all concerned. Most of the issue reports associated with the top 10 privacy requirements were bug, followed by feature issue type. This implies that Chrome had not properly implemented the functionalities related to providing users the way to execute their individual rights, obtaining and managing user consent and preferences and providing notice to users.}

\newtext{Four categories including user participation, notice, security and user desirability were covered by the top 10 privacy requirements in Moodle. The data subjects and consent subcategories in the notice and user desirability categories were covered by this set of privacy requirements respectively. Unlike Chrome, the majority of the issue reports associated with the top 10 privacy requirements were feature request, followed by bug type. Moodle focused on implementing new features that allow users to execute their individual rights and inform users of relevant privacy-related information. In addition, Moodle also emphasised on implementing security measures to protect personal data.}


\subsection{The treatment of privacy and non-privacy issues}

We investigate if privacy issues were treated differently from non-privacy issues in Chrome and Moodle. We focus on observing two kinds of treatments: the time it took to resolve an issue and the number of comments associated with the issue. The former reflects how fast an issue was resolved while the latter indicates the attention and engagement of the project team to the issue. We randomly sampled the dataset we built earlier using a 95\% confidence level with a confidence interval of 5\footnote{https://www.surveysystem.com/sscalc.htm} to obtain 269 privacy issues from Chrome and 213 from Moodle. Applying the same sampling scheme, we randomly selected 382 non-privacy issues from Chrome and 380 from Moodle - these issues were not tagged as privacy in the ``component'' field. Note that the resolution time is calculated from the number of days between reported date and the date when the issue was flagged as being resolved.

\begin{table}
	\centering
	\caption{Results of the Wilcoxon rank-sum test: non-privacy vs. privacy issues}
	\label{tab:ranksum}
	\resizebox{8.5cm}{!}{
		\begin{tabular}{l l l l l}
			\toprule
			\textbf{Project} & \textbf{Attribute} & \textbf{One-sided tail} & \textbf{p-value} & \textbf{Effect size}\\
			\midrule
			Google Chrome & Resolution time & Less & $<$0.001 & 0.578 \\
			Google Chrome & \#Comments & Less & $<$0.001 & 0.691 \\
			Moodle & Resolution time & Greater & $<$0.001 & 0.609 \\
			Moodle & \#Comments & Greater & $<$0.001 & 0.604\\
			\bottomrule
		\end{tabular}%
	}	
\end{table}

We employ the Wilcoxon rank-sum test (also known as Mann-Whitney U test), a non-parametric hypothesis test which compares the difference between two independent observations \cite{Wild1997}. We performed two tests between privacy and non-privacy samples, one for the resolution time and the other for the number of comments. The results (see Table \ref{tab:ranksum}) show that the resolution time and the number of comments are statistically significantly (\textit{p-value $\leq$ 0.001}) different between privacy and non-privacy issues in both Chrome and Moodle with effect size greater than 0.5 in all cases.

We also compare the median rank of the two samples using one-tailed test. Our results show that privacy issues were resolved more quickly and attracted less comments than non-privacy issues in Moodle (see Table \ref{tab:ranksum}). On the other hand, it took longer to resolve privacy issues than non-privacy issues in Chrome. Also, privacy issues in Chrome tend to attract more discussion than non-privacy issues. 


\newtext{We observed the following patterns when we went through the comments in non-privacy issue reports in Chrome: (i) the issue reports were assigned to relevant contributors and got resolved without any discussion (e.g. Issue 142322\footnote{https://bugs.chromium.org/p/chromium/issues/detail?id=142322)}, (ii) the contributors asked if fixing those issue reports did not affect other parts without discussing on how to fix the issues (e.g. Issue 914196\footnote{https://bugs.chromium.org/p/chromium/issues/detail?id=914196}) and (iii) the contributors discussed on workarounds meaning that they knew how to fix the issues but direct method could not be used (e.g. Issue 1082077\footnote{https://bugs.chromium.org/p/chromium/issues/detail?id=1082077}). These patterns show that the contributors had less discussion with regard to identifying root causes or solutions of the issues. Based on these findings, the contributors in Chrome tended to be confident and had more experience in resolving non-privacy issues.} 

By contrast, privacy issues in Chrome attracted more discussions since the contributors were uncertain about the issues and their affected components in the system. We observed five examples that the contributors commented in those privacy issues as follows: (i) the contributors did not know what the affected components reported in the issues do (e.g. issue 345741); (ii) the contributors could not identify the causes of issues; (iii) the contributors required time and effort to come up with potential solutions; (iv) the contributors needed to assess the difficulties of the issues and their resolutions; and (v) the contributors did not know whom to assign the work. These reasons also led to longer time to resolve the privacy issues in Chrome. In addition, the Chrome project does not have a well-defined process that specifically handles privacy. Hence, the contributors need to ensure that fixing privacy issues will not create another problem in different components. Thus, resolving privacy issues attracted a lot of discussions, leading to longer resolution time.

On the other hand, privacy issues in Moodle were resolved more quickly and attracted less comments than non-privacy issues. We observe that the privacy issues were well reported and clearly explained in Moodle. Moodle contributors were familiar with privacy-related functionalities and relevant system components. In addition, Moodle has a clearly defined infrastructure to handle privacy and privacy compliance in the system (e.g. privacy API \cite{Moodle2019} and GDPR for plugin developers \cite{Nicols2018}). This infrastructure includes a number of components that support privacy-related functionalities and several key individual rights in GDPR (e.g. accessing to personal data and requesting for deletion). When there is a privacy-related bug or new feature request, the contributors can consult the privacy API documentation and identify the components that they must fix or implement. Hence, the privacy issues took less time and attracted less comments in Moodle.

\section{Threats to validity} \label{sec:threats}


Our study involved subjective judgements. We have applied several strategies to mitigate this threat such as using multiple coders (who are the authors of the paper), applying IRR assessments, organising training sessions and disagreement resolution meetings. A legal expert could have extended the view of the human coders in interpreting legal regulations. However, we note that all the coders had attended a training course on privacy regulations (including GDPR). This has enhanced our interpretation of privacy regulations from a legal perspective, thus minimised this risk. In addition, our study was built upon well-founded processes and theories in previous work such as Grounded Theory \cite{Glaser2017} and GBRAM \cite{Antn2004}. We acknowledge that other contemporary methods could be used to extract requirements from legal texts (e.g. \cite{Breaux2006}, \cite{Ghanavati2009}). We also used relevant statistical measures and techniques to ensure that our findings are statistically significant. \\
\indent We are aware that there are other privacy protection laws and regulations applied around the world. However, most of them share many commonalities with the GDPR and ISO/IEC 29100 as confirmed by Thailand PDPA and APEC privacy framework in this study. 
In fact, GDPR is one of the most comprehensive data protection regulations \cite{Linden2020}, \cite{Tsohou2020}. It was also used as a benchmark for other countries to develop data protection regulations such as Japan, South Korea and Thailand \cite{Torre}, \cite{Laboris2019}. Hence, although the principles or rights in other laws and regulations are slightly different due to variations in each country/city, they share many commonalities with GDPR (e.g. see the comparison between GDPR and CCPA in \cite{DataPrivacyManager}). ISO/IEC 29100 has also been used to develop organisational and technical privacy controls in many information and communication systems \cite{PECB2015}. Therefore, we found that GDPR and ISO/IEC 29100 together are the most comprehensive, thus our taxonomy can generalise to other privacy regulations and standards. We acknowledge that future research could involve investigating country specific privacy regulations, and extending our taxonomy accordingly.  \\
\indent It is noted that the taxonomy does not address all the levels of abstractions of privacy requirements. The privacy requirements can be refined into a subset of other requirements. For example, the notion of consent as a privacy requirement can be refined into multiple requirements such as consent methods and properties of consent management platform. This depends on business requirements, organisational processes and software development teams which limit the applicability of our taxonomy. Also, the future amendments to the regulations and frameworks may require updates of our taxonomy. \\
\indent \newtext{In a modern software development, requirements collected during the requirement elicitation phase can be specified and recorded in issue-tracking systems (such as JIRA). However, we acknowledge that issue reports may not capture all requirements of a software application, e.g. requirements associated with the software's core architecture. Our methodology can be extended to map those requirements into the privacy taxonomy. Future work would involve combining our approach with other sources of software requirements to provide a more complete view of how a software application addresses the privacy requirements in the taxonomy.}

Finally, we performed the mining and classification of issue reports for Chrome and Moodle. These are two large and widely-used software systems that have strong emphasis on privacy concerns. However, we acknowledge that our datasets may not be representative of other software applications. Further investigation is required to explore other projects in different domains (e.g. e-health software systems and mobile applications). However, we note that building this dataset on two projects alone required substantial effort and highly thorough processes (278 person-hours).

\section{Conclusion and future work} \label{sec:conclusion}

In this paper, we have developed a comprehensive taxonomy of privacy requirements based on GDPR, ISO/IEC 29100, Thailand PDPA and APEC privacy framework. Our approach is built upon a content analysis process, adapted from the GBRAM which are generic and applicable to different regulations and privacy standards. We performed reliability assessments and disagreement resolution in the process to ensure that our taxonomy is reliably constructed. Our taxonomy consists of 71 privacy requirements grouped in 7 privacy goal categories. Since the studied regulations and frameworks are not specific to any software types, our taxonomy is generally applicable to a wide range of software applications.

We have also performed a study on how two large projects (Chrome and Moodle) address those privacy requirements in our taxonomy. To do so, we mined the issue reports recorded in those projects and collected 1,374 privacy-related issues. We then classified these issues into our taxonomy through a process which involved multiple coders and the use of IRR assessments and disagreement resolution. We found that the privacy requirements in the user participation category were covered in a majority of the issues. We also found that the time taken to resolve privacy-related issues and the degree of developers' engagement on them were also statistically significantly different from those of non-privacy issues.

Our work lays several important foundations for future research in this area. The systematic method performed in the work enables future research to be conducted on other data protection and privacy regulations and frameworks. The taxonomy can act as a reference for the research community to discuss and expand. We plan to investigate other privacy regulations and policies and extend our taxonomy, if necessary. The requirements in our taxonomy are written in natural language and structured into templates. Although we believe that this is the most intuitive form to developers, future work could explore other alternative forms such as semantic frame-based representation \cite{Bhatia2019}. We have manually derived requirements in this study as it is essential to examine structure of statements and how privacy requirements are expressed in different regulations and frameworks. However, the requirements extraction process can be automated using NLP techniques (e.g. \cite{Zeni2015} and \cite{Sleimi2018}).

Future work also involves exploring how issue reports in other projects (e.g. health and mobile applications) attend to the requirements in our taxonomy. This also includes other issues that are not tagged as privacy-related. In addition, software developers would be able to participate in validating the use of taxonomy and their understanding towards the derived privacy requirements. Legal experts could also be involved to help interpret legal perspective of the taxonomy. Furthermore, we plan to develop tool support to automate the privacy requirements identification and classification tasks when users report issues in ITS. Other potential future work includes the investigation of the use of the taxonomy with semantic web technologies to facilitate computational and regulatory risk analysis purposes. The investigation of other forms of traceability relationships such as the traceability between code and issue reports or code and privacy requirements can also be further studied.

\balance

\bibliographystyle{elsarticle-num-names}
\bibliography{privacy-requirements,ICSE2020_kookai_ref}

\end{document}